# The Discretised Lognormal and Hooked Power Law Distributions for Complete Citation Data: Best Options for Modelling and Regression[1]

Mike Thelwall, Statistical Cybermetrics Research Group, University of Wolverhampton, UK.

Identifying the statistical distribution that best fits citation data is important to allow robust and powerful quantitative analyses. Whilst previous studies have suggested that both the hooked power law and discretised lognormal distributions fit better than the power law and negative binomial distributions, no comparisons so far have covered all articles within a discipline, including those that are uncited. Based on an analysis of 26 different Scopus subject areas in seven different years, this article reports comparisons of the discretised lognormal and the hooked power law with citation data, adding 1 to citation counts in order to include zeros. The hooked power law fits better in two thirds of the subject/year combinations tested for journal articles that are at least three years old, including most medical, life and natural sciences, and for virtually all subject areas for younger articles. Conversely, the discretised lognormal tends to fit best for arts, humanities, social science and engineering fields. The difference between the fits of the distributions is mostly small, however, and so either could reasonably be used for modelling citation data. For regression analyses, however, the best option is to use ordinary least squares regression applied to the natural logarithm of citation counts plus one, especially for sets of younger articles, because of the increased precision of the parameters.

## 1. Introduction

The citation impact of sets of articles from journals (Chandy & Williams, 1994), researchers (Meho & Yang, 2007), research groups (van Raan, 2006), departments (Oppenheim, 1995), universities (Charlton & Andras, 2007) or even countries (Braun, Glänzel, & Schubert, 1985) are often compared with quantitative indicators on the basis that citations tend to reflect scientific impact. In addition, sets of articles with different properties are also sometimes analysed with the aid of citation counts, such as to test whether open access articles tend to be more frequently cited (Harnad & Brody, 2004; McCabe & Snyder, 2015) or whether collaboration tends to increase citations (Gazni & Didegah, 2011; Glänzel, Schubert, & Czerwon, 1999). These comparisons often employ standard indicators, such as the h-index (Hirsch, 2005) or field normalised citation counts (Waltman, van Eck, van Leeuwen, Visser, & van Raan, 2011). If part of a formal evaluation, then these indicators may be used to inform qualitative judgements. For more theoretical reasons, citation counts are sometimes analysed using statistical regression, where the independent variables are factors to be tested for a relationship with research impact, such as the number or nationality of the authors (Didegah & Thelwall, 2013; Onodera & Yoshikane, 2015; Yu, Yu, Li, & Wang, 2014). For both of these purposes, it is essential to understand the broad properties of sets of citation counts so that the indicators developed, and the regression approaches used, can be as powerful and appropriate as possible. This is particularly important because citation counts are known to be highly skewed and so many statistical techniques, including the sample mean, are not appropriate for them.

2There have been many studies of citation count distributions since the early realisation that they were highly skewed, with a small number of articles attracting very high citation counts (de Solla Price, 1965). This skewed nature was thought to be due to preferential attachment processes in science (the Matthew effect), with articles attracting citations at least partly because they had already been cited (de Solla Price, 1976; Merton, 1968). This process is possible because researchers can find cited articles from other articles' reference lists, being cited can grant prestige, and modern digital libraries, such as Google Scholar, tend to list more cited articles above less cited articles. Nevertheless, articles attract citations much more rapidly than accounted for by the publication lifecycle and so preferential attachment cannot fully explain the pattern of growth in citations because few authors can cite an article using knowledge about how may citations it will have attracted when their work is published. To investigate this, one study has found evidence from physics that interest in an article decays exponentially over time (Eom & Fortunato, 2011).

Several studies have shown that citation counts tend to follow a power law distribution (or variants: van Raan, 2001) quite well, at least if articles with few citations are excluded (Clauset, Shalizi, & Newman, 2009; Garanina, & Romanovsky, 2015; Redner, 1998). This is sometimes described as fitting the tail of the distribution. The hooked/shifted power has been show to fit better than the power law and about as well as the discretised lognormal distribution for citations to papers from 12 American Physics Society journals if articles with few citations are excluded (Eom & Fortunato, 2011). Some regression analyses have the negative binomial distribution instead (e.g., Didegah & Thelwall, 2013; Hanssen & Jørgensen, 2015; Onodera & Yoshikane, 2015), on the basis that it is for discrete data and can cope with highly skewed data. It does not fit citation distributions as well as the discretised lognormal, however (Low, Thelwall, & Wilson, 2015), because of the heavy tailed nature of sets of citation counts (i.e., relatively many very high values within the data). Conversely, the Yule-Simon distribution, which is essentially a discrete version of the power law based upon assumptions about preferential attachment, seems to fit the tail of citation count distributions well (Brzezinski, 2015). Nevertheless, it unlikely to fit citation distributions well if zeros are included and it is shifted by 1 to allow zeros, because it is a strictly decreasing function and in some fields the mode is not zero (e.g., Developmental Biology: Radicchi, Fortunato, & Castellano, 2008).

For articles from a single subject and year, if uncited articles (only) are excluded, then the discretised lognormal (Evans, Hopkins, & Kaube, 2012; Radicchi, Fortunato, & Castellano, 2008) and hooked power law (Pennock, Flake, Lawrence, Glover, & Giles, 2002) (see below for descriptions of the distributions) fit substantially better than the power law distribution (Thelwall & Wilson, 2014a) and there do not seem to be any serious alternatives (excluding those with unstable parameters: Low, Thelwall, & Wilson, 2015). Uncited articles are typically removed when fitting most distributions because some of them, including the power law and discretised lognormal, are usually implemented in a way that excludes zeros, although logarithmic binning is a way of avoiding this problem (Evans, Hopkins, & Kaube, 2012). The omission of uncited articles is a problem since they are important for any full analysis of groups of articles. Hence, approaches are also needed to model the full range of citation counts.

One article has previously addressed this issue by comparing negative binomial and lognormal regression models for citation count data in a way that includes uncited articles, using 1337 journal articles published between 2001 and 2010 matching a Scopus title search for "knowledge management", and using as independent variables the number of years since publication and the number of references in the article. It also analysed a data set of articles from the online Information Research journal between 2001 and 2011, and using as independent variables the number of website views, Mendeley readers, and years since publication (Ajiferuke & Famoye, 2015). The negative binomial regression model was found to fit better than the discretised lognormal model but in both cases the data sets are relatively small, and the use of the publication year as an independent variable for a data set with multiple years is problematic because the relationship between publication year and citation counts is not simple (Adams, 2005; Eom & Fortunato, 2011) and hence may not be modelled well by regression. A previous study using simulations had shown that negative binomial regression had a tendency to identify non-existent relationships at a rate above the significance level set, showing that conclusions drawn from negative binomial regression are unsafe (Thelwall & Wilson, 2014b). Whilst this conclusion was not confirmed by the analysis of Information Research articles and knowledge management articles (Ajiferuke & Famoye, 2015), the number of dependant variable tested was too small and the nature of the datasets tested too restricted to give convincing evidence and so the use of negative binomial regression for citation data remains problematic.

This article uses a simple approach to model uncited articles with distributions that do not allow zeros: adding 1 to all citation counts before fitting a model. This simple transformation, which is a common way of dealing with zeros in a dataset that needs a log transformation (O'Hara & Kotze, 2010), allows the discretised lognormal distribution to be fitted to the full range of data and allows it to be compared against the main current alternative, the hooked power law. This transformation could perhaps be justified on the theoretical grounds that each article announces itself by its existence and is therefore a kind of self-citation. If data naturally fits the negative binomial distribution, however, then it is preferable to use negative binomial regression than to log transform the data before using regression (O'Hara & Kotze, 2010). This article compares the discretised lognormal and hooked power law distributions for the transformed data and also analyses fitting the normal distribution to the log transformed data as an alternative.

## 2. The discretised lognormal and hooked power law distributions

Many studies have proposed distributions that attempt to improve on the power law for fitting various data sets (e.g., Dorogovtsev, Mendes, & Samukhin, 2000; Levene, Fenner, Loizou, & Wheeldon, 2002). The hooked power law, as used for web page link distributions (Pennock, Flake, Lawrence, Glover, & Giles, 2002), is based upon the assumption that articles get cited through two separate processes. For the first process, citations are generated at random (although in practice they may be attracted primarily by the intrinsic qualities of articles). For the other process, new citations are driven by existing citations so that the probability that an article attracts a new citation is only related to the number of citations that it has received. Combining these two processes then the probability that an article attracts $k$ citations is proportional to $1/(B+k)^\alpha$. The two parameters affect both processes and setting the offset parameter B to zero generates a pure power law.

The lognormal distribution (Clauset, Shalizi, & Newman, 2009) is a (continuous) probability distribution that fits distributions when the natural logarithm of the data follows a normal distribution. Because the natural logarithm can only be calculated for positive numbers, data with negative numbers and zero cannot be used for fitting the distribution. The lognormal distribution has a mean and standard deviation parameters, µ and σ, but the expected mean of the distribution is related to both: $e^{\mu+\sigma^2/2}$. Since this is a continuous probability distribution, when applied to discrete data, the same distribution formula can be used (i.e., the probability density function is $f(x) = \frac{1}{x\sigma\sqrt{2\mu}} e^{-\frac{(\ln(x-\mu))^2}{2\sigma^2}}$ is treated as a point mass probability function) as long as a correction is made to ensure that the sum of the probabilities is 1. In other words, the probability of a citation count of $n$ is $f(n)/\sum_{n=1}^{\infty} f(n)$, if uncited articles are excluded. As discussed below, when citation counts were included, an offset of 1 was used so that the probability of a paper receiving $n$ citations was modelled as $f(n+1)/\sum_{n=0}^{\infty} f(n+1)$. Although there are alternative methods of converting the continuous lognormal into a discrete distribution, such as integrating the probability density function over a unit interval around a given point (e.g., the discrete probability of 1 could reasonably be equated with the continuous probability of either of the intervals (0.5,1.5) or (0,1]), the method chosen here seems reasonable in the absence of convincing evidence that an alternative is better. This decision particularly affects the probability of citation counts when the probability distribution is rapidly changing, such as for uncited articles.

## 3. Research questions

The following research questions drive the investigation. Although the first research question has been previously answered, as discussed above, the impact of time on the answer has not been investigated and so it is included for completeness. For regression analysis of citation data it is important not only to assess which distribution fits the data best but also how stable the fitted parameters are. This is because regression is likely to be more powerful if the parameters can be estimated more precisely.

1. Which of the discretised lognormal distribution and the hooked power law best fits citation data, when uncited articles are excluded?
2. Which of the discretised lognormal distribution and the hooked power law best fits citation data, when uncited articles are included?
3. For the discretised lognormal distribution, how do the fitted parameters vary by field and year?
4. If which of the hooked power law and the discretised lognormal give the most robust parameters when fitted to the data? Are these parameters more robust than the parameters from the normal distribution fitted to the log transformed data (i.e., log(1+x))?
5. How do the answers to the above questions vary by field and by year?

## 4. Data and Methods

The data is the citation counts to articles from 26 Scopus subject areas and seven years, re-used from two previous articles (Thelwall & Fairclough, 2015a; Fairclough & Thelwall, 2015b), and collected from April 15 to May 11, 2015 (see Table 3). This data is suited to addressing the research questions because it covers a wide range of different subject areas and a range of different years. It includes 911,971 journal articles, 610,626 of which had received at least one citation. Data from 2015 was included despite its low citation counts. This is because citation analyses is often used to inform policy and decision makers and, in this context, it is important to use the most current information available and so findings about partial data from the current year may be useful.

Since uncited data cannot be fit to the standard discretised lognormal because zeros are not permitted, 1 was added to all citation counts in order to enable all data to be used in the fitting process when uncited articles were included. A transformation by adding 1 is a standard technique for dealing with zeros in statistical calculations that would otherwise be impossible or meaningless, such as the geometric mean. It has also been used in regression for citation data [Thelwall & Wilson, 2014b) and has been implicitly used in geometric mean calculations for citations (Fairclough & Thelwall, 2015a; Thelwall, in press). It is not needed for the hooked power law distribution but it does not change the fit of the model to the data (it just decreases the value of the fitted B parameter by 1) and so, for simplicity, the hooked power laws were also fitted to the same transformed data.

The discretised lognormal distribution was fitted to the data using the R poweRlaw package and the hooked power law was fitted using R code modified from a previous paper (Thelwall & Wilson, 2014a), which uses a gradient descent method to fit the α and *B* parameters of the model. Normal distributions were also fitted to the data after adding 1 to the citation counts and taking the natural log. The R MASS package was used for this, which uses the standard formulae for the normal distribution parameters. All code, data and results are available in the online appendix. The hooked power law and discretised lognormal fits were compared using the Vuong test for non-nested models (Vuong, 1989).

The robustness of parameters for distributions can be assessed, in part, by examining the extent to which they change for the same subject area over the years. Both unchanging parameters and smooth parameter changes over time would be consistent with the estimates being reasonably reliable, whereas substantial differences between years indicates either that the parameter estimates are not very accurate or that they are too erratic for regression purposes. A previous study for simulated data without zeros (Thelwall & Wilson, 2014a) has found that the hooked power law and discretised lognormal distribution parameter estimates are not very precise but it is important to check this with zeros included for real data, and also to compare with the standard normal distribution for the transformed data. A simple way to assess the extent to which the parameters are either stable or change smoothly over time is to compare the rank order of the parameters between successive years. The more precise the parameter estimates are, the more likely the rank order of the parameters between subjects would be the same between consecutive years. Hence, the Spearman correlation between the parameter values in successive years was used as the primary robustness check, backed up by a visual inspection of the changes in parameter values over time.

## 5. Results

For most years citation counts in some subject areas were fit best by a hooked power law and in other areas best by a lognormal distribution, whether or not zeros were excluded (Table 1) or included (Table 2). When uncited articles are excluded (Table 1), the hooked power law fits better more often than when they are included (Table 2). The data fits the discretised lognormal statistically significantly better than the hooked power law in only two cases out of 182, which is broadly consistent with the hooked power law being a universally better fit when uncited articles are excluded (Table 1). Nevertheless, the differences in many cases are not statistically significant despite the relatively large sample sizes in almost all cases, suggesting that the differences in fit between the two distributions tends not to be substantial.

When uncited articles are included and 1 is added to the citation counts before fitting (Table 2), the lognormal distribution tends to fit best a third of the time for articles that are at least 2 years old, but the hooked power law fits best more often for younger articles and for two thirds of the subject areas for older articles. Excluding the partial year of 2015, the subject areas are also reasonably consistent in the distribution that fits best. This is despite the data for each year being completely different sets of articles, although some of the citing articles would be the same. This suggests that neither distribution is universally most suitable for citations in all subject areas, when uncited articles are included. Instead, some subject areas seem to fit one of the two distributions naturally better than the other. A factor that might be relevant for this is the extent to which a category is homogeneous for its subject area.

**Table 1**. A comparison of the hooked power law and the discretised lognormal distribution for Scopus journal articles in 26 subjects and 7 publication years, *excluding* uncited articles.

| Subject* | 2009 | 2010 | 2011 | 2012 | 2013 | 2014 | 2015 | Total art. |
|---|---|---|---|---|---|---|---|---|
| Animal Science and Zoology | **Hook** | **Hook** | Hook | **Hook** | **Hook** | Hook | Hook | 34539 |
| Language and Linguistics | **Hook** | Hook | Hook | L | Hook | L | Hook | 14150 |
| Biochemistry | **Hook** | **Hook** | Hook | Hook | **L** | Hook | Hook | 39448 |
| Business and International Management | L | L | Hook | Hook | L | Hook | Hook | 17214 |
| Catalysis | L | L | L | L | L | Hook | Hook | 47763 |
| Electrochemistry | **Hook** | Hook | Hook | Hook | L | **Hook** | Hook | 47285 |
| Computational Theory and Mathematics | **Hook** | Hook | Hook | Hook | Hook | **Hook** | Hook | 18307 |
| Management Science & Operations Res. | **Hook** | **Hook** | **Hook** | **Hook** | Hook | L | Hook | 25800 |
| Computers in Earth Sciences | **Hook** | **Hook** | **Hook** | **Hook** | Hook | Hook | Hook | 7164 |
| Finance | Hook | L | Hook | Hook | Hook | L | Hook | 25359 |
| Fuel Technology | **Hook** | **Hook** | **Hook** | **Hook** | **Hook** | **Hook** | Hook | 19461 |
| Automotive Engineering | L | L | Hook | Hook | Hook | Hook | Hook | 11688 |
| Ecology | **Hook** | **Hook** | **Hook** | **Hook** | Hook | Hook | Hook | 35646 |
| Immunology | Hook | Hook | Hook | Hook | L | L | Hook | 35227 |
| Ceramics and Composites | **Hook** | **Hook** | Hook | Hook | **Hook** | Hook | Hook | 27867 |
| Analysis | Hook | **Hook** | Hook | L | Hook | Hook | L | 32077 |
| Anesthesiology and Pain Medicine | **Hook** | **Hook** | **Hook** | **Hook** | **Hook** | Hook | Hook | 22879 |
| Biological Psychiatry | Hook | L | Hook | L | L | Hook | Hook | 17543 |
| Assessment and Diagnosis | L | **L** | L | Hook | L | L | Hook | 1181 |
| Pharmaceutical Science | **Hook** | **Hook** | **Hook** | **Hook** | **Hook** | Hook | Hook | 28341 |
| Astronomy and Astrophysics | **Hook** | **Hook** | **Hook** | **Hook** | **Hook** | Hook | Hook | 37056 |
| Clinical Psychology | **Hook** | **Hook** | **Hook** | **Hook** | **Hook** | Hook | L | 33181 |
| Development | Hook | Hook | **Hook** | **Hook** | **Hook** | Hook | Hook | 19771 |
| Food Animals | **Hook** | Hook | Hook | **Hook** | Hook | Hook | **Hook** | 6977 |
| Orthodontics | **Hook** | Hook | Hook | Hook | **Hook** | Hook | L | 3575 |
| Complementary and Manual Therapy | Hook | **Hook** | Hook | Hook | Hook | Hook | **Hook** | 1127 |
| **Percentage with lognormal fitting best** | 15% | 23% | 8% | 15% | 27% | 19% | 12% | 610626 |

*Hook: hooked power law has the better fit; L: lognormal has the better fit; bold and underlined: difference between the two fits is statistically significant at the α=0.05 level according to the Vuong test. See online appendix for fit statistics.

**Table 2**. A comparison of the hooked power law and the discretised lognormal distribution for Scopus journal articles in 26 subjects and 7 publication years, including all uncited articles, and with 1 added to all citation counts.

| Subject | 2009 | 2010 | 2011 | 2012 | 2013 | 2014 | 2015 | Total art. |
|---|---|---|---|---|---|---|---|---|
| Animal Science and Zoology | **Hook** | **Hook** | **Hook** | **Hook** | **Hook** | Hook | **Hook** | 56008 |
| Language and Linguistics | **L** | **L** | **L** | **L** | **Hook** | Hook | Hook | 34787 |
| Biochemistry | **Hook** | **Hook** | **Hook** | **Hook** | Hook | L | **Hook** | 56576 |
| Business and International Management | **L** | **L** | **L** | **L** | L | Hook | **Hook** | 32879 |
| Catalysis | **L** | **L** | **L** | **L** | **L** | Hook | **Hook** | 58377 |
| Electrochemistry | **Hook** | **Hook** | Hook | L | L | Hook | **Hook** | 57698 |
| Computational Theory and Mathematics | **Hook** | Hook | Hook | **Hook** | Hook | **Hook** | Hook | 27850 |
| Management Science & Operations Res. | **Hook** | **Hook** | **Hook** | **Hook** | Hook | L | Hook | 36736 |
| Computers in Earth Sciences | **Hook** | **Hook** | **Hook** | Hook | Hook | Hook | Hook | 9598 |
| Finance | **L** | **L** | L | Hook | Hook | Hook | Hook | 41493 |
| Fuel Technology | **L** | **L** | **L** | Hook | L | L | **Hook** | 32202 |
| Automotive Engineering | **L** | **L** | **L** | L | L | Hook | Hook | 19891 |
| Ecology | **Hook** | **Hook** | **Hook** | **Hook** | **Hook** | Hook | **Hook** | 51862 |
| Immunology | **Hook** | **Hook** | **Hook** | **Hook** | Hook | Hook | Hook | 45847 |
| Ceramics and Composites | **Hook** | **Hook** | **Hook** | **Hook** | **Hook** | Hook | Hook | 40963 |
| Analysis | **Hook** | **Hook** | **Hook** | Hook | L | Hook | Hook | 52890 |
| Anesthesiology and Pain Medicine | **L** | **L** | Hook | **L** | **L** | L | Hook | 35005 |
| Biological Psychiatry | Hook | Hook | Hook | Hook | **L** | Hook | Hook | 22040 |
| Assessment and Diagnosis | L | L | L | **L** | L | Hook | Hook | 2690 |
| Pharmaceutical Science | **Hook** | **Hook** | **Hook** | Hook | L | Hook | **Hook** | 46750 |
| Astronomy and Astrophysics | **Hook** | **Hook** | **Hook** | **Hook** | **Hook** | Hook | **Hook** | 48196 |
| Clinical Psychology | **Hook** | **Hook** | **Hook** | **Hook** | **Hook** | Hook | **Hook** | 49920 |
| Development | **L** | **L** | L | L | Hook | Hook | **Hook** | 33862 |
| Food Animals | **Hook** | **Hook** | Hook | Hook | **Hook** | Hook | Hook | 9872 |
| Orthodontics | **Hook** | **Hook** | **Hook** | **Hook** | **Hook** | Hook | Hook | 5947 |
| Complementary and Manual Therapy | Hook | **Hook** | Hook | Hook | Hook | Hook | Hook | 2032 |
| Lognormal percentage | 35% | 35% | 31% | 31% | 38% | 15% | 0% | 911971 |

*Hook: hooked power law has the better fit; L: lognormal has the better fit; bold and underlined: difference between the two fits is statistically significant at the α=0.05 level according to the Vuong test. See online appendix for fit statistics.

For articles that are at least three years old, all four methods seem to generate results that are stable over time, although there is substantially less stability for more recent years (Figure 1). The normal distribution mean parameter is the most stable (has the highest spearman correlation between years) in five of the six pairs of years checked, and is close to being the most stable in the remaining case and so the mean parameter of the normal distribution seems to be the most stable of all parameters. Although the normal distribution standard deviation parameter is the least stable parameter for the oldest three years, the difference in stability is not large and it is substantially more stable than the parameters from all other distributions for the most recent data. Taking into account that the mean parameter is more important than the standard deviation parameter for citation distributions (because it can be used for an average impact indicator) the continuous normal distribution fitted to log transformed citation counts is the best choice for all citation data when distribution parameters need to be estimated, despite the hooked power law providing a better overall fit Table 2, and the only realistic choice (out of the three tested here) for articles that are less than a year old.

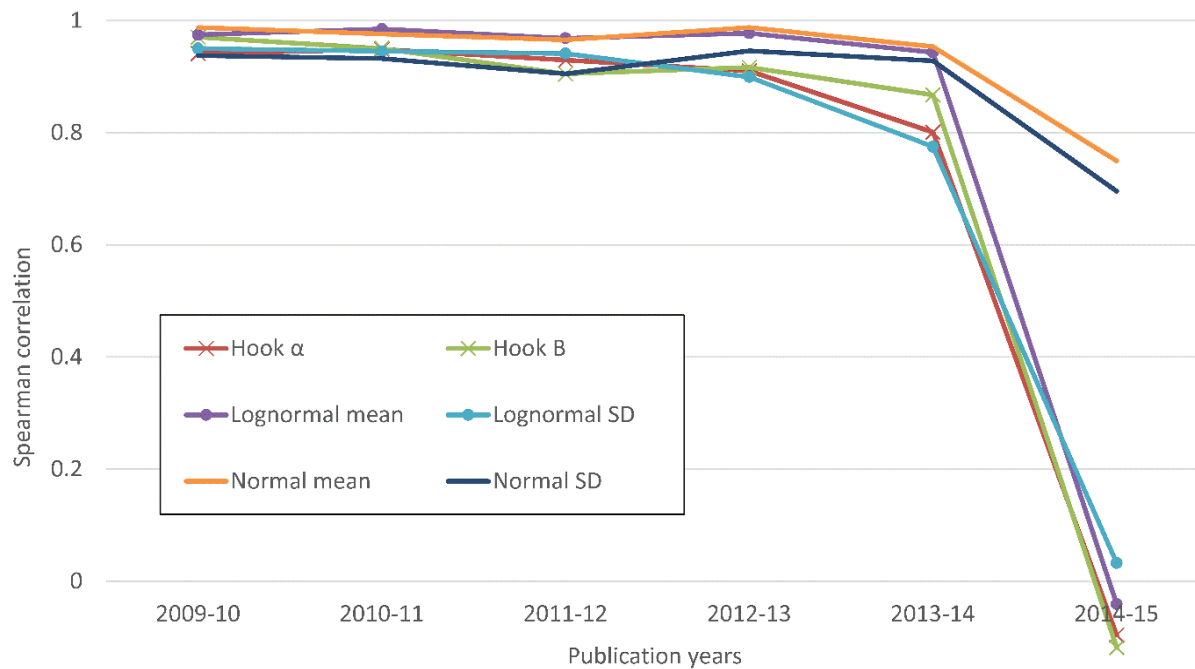

**Figure 1**. The Spearman rank order correlation for the parameters from the three distributions in successive years, including all uncited articles, and with 1 added to all citation counts (i.e., the Table 2 data).

## 6. Discussion

The study has a number of limitations. Not all subject areas were investigated and so it may be that the results do not apply to some areas. This seems to be particularly likely for fields that do not use citations in the normative way of science, such as the arts and humanities (Hellqvist, 2010; Ngah, & Goi, 1997). The findings are also dependent upon the breadth of the subject areas investigated. It is possible that broader collections of articles from multiple fields may have different properties to those found here. Biases in the selection of journals for each subject category may also affect the results by omitting articles that do not fit the distribution, for example by tending to have lower citation counts. This is perhaps particularly relevant to articles that are not in English (Archambault, Vignola-Gagne, Côté, Larivière, & Gingras, 2006; Hicks, 1999; van Leeuwen, Moed, Tijssen, Visser, & van Raan, 2001). The patterns may also not hold for older years, although this seems unlikely because six years seems to be enough for most articles to attract most citations. Nevertheless, the distribution may change if only a few highly cited old articles continue to attract citations whilst the rest are ignored, for example. Finally, there may be other distributions that fit the data better than any distribution tested so far, and there may be other ways of fitting the current distributions. For example, it would have been possible to add more than 1 to the data before fitting the lognormal distribution, which would have the effect of truncating its left hand side.

This discussion focuses on the data type that is the focus of this article: the complete sets of articles, including those without citations. Ignoring the most recent two years, there do not seem to be strong broad disciplinary patterns in the model that fits the data best when uncited articles are included. Within the social sciences, arts and humanities, the discretised lognormal might be the best overall, however. It tends to fit best in three cases (Language and Linguistics; Business and International Management; Development) but the hooked power law is better for another (Management Science & Operations Research) and one case is mixed (Finance). The engineering and technology fields (Fuel Technology; Catalysis; Automotive Engineering) also favour the discretised lognormal. For health the discretised lognormal fits best in only two cases (Anesthesiology and Pain Medicine; Assessment and Diagnosis) but the hooked power law fits best in six (Biological Psychiatry; Pharmaceutical Science; Immunology; Clinical Psychology; Orthodontics; Complementary and Manual Therapy) and so the hooked power overall seems to be the most appropriate (agreeing with a previous study of 45 medical fields from a single year: Thelwall & Wilson, in press), as it is for the remaining (natural and life) science fields. Overall, however, the results broadly fit the pattern that the hooked power law fits best in the scientific areas where citations are regarded as more important (life, natural and health sciences) but the lognormal fits best in areas where citations are arguably less important (engineering and technology, social sciences, arts and humanities) and for which citations may be used in a different way.

The correlations (Figure 1) point to the normal distribution producing the most stable parameters. The parameter estimates for the discretised lognormal distribution (Figures 2 and 3) are mostly stable, generating smooth lines as the mean and standard deviation parameters both tend to increase over time. Nevertheless, they are erratic for some of the subject areas and for the most recent year. The hooked power law parameters are much less stable over time (Figures 4 and 5) but the normal distribution parameters changes are mostly smooth (Figures 6 and 7). A gradual decrease is expected, at least for the mean, due to older articles tending to be more cited than younger articles. The subject areas of Assessment and Diagnosis and Language and Linguistics have the most unstable behaviour over time for the discretised lognormal distribution. This is not a by-product of a small sample size for Language and Linguistics because it includes more than 7800 articles for each year except 2015 (4855 articles), although Assessment and Diagnosis includes between 382 and 486 articles before 2015 (110 articles). Strangely, both are relatively stable for the hooked power law before 2015 and so their instability is specific to the discretised lognormal distribution, despite fitting it better than the hooked power law. Thus, the degree of fit does not determine the robustness of the parameter estimates, as previously noted for a comparison of different distributions for citation data (Low, Thelwall, & Wilson, 2015). For the hooked power law, Biological Psychiatry probably has the most unstable behaviour over time for both parameters despite relatively large sample sizes (above about 2800 before 2015) and its parameters are relatively stable for the discretised lognormal distribution. Hence the unstable parameter estimates appear to be due to a characteristic of the subject area rather than the sample size.

A root cause of the potential for instability over time is that the parameters for both the discretised lognormal and hooked power law distributions do not model the mean separately from the standard deviation, in contrast to the normal distribution. For the discretised lognormal distribution, the expected mean of the distribution is dependent upon both the mean parameter and the standard deviation parameter and the same is true for the hooked power law α and B parameters. This issue is related to the difficulty with generating precise parameter estimates for both distributions, which has been shown before (using bootstrapping) when uncited articles are excluded (Thelwall & Wilson, 2014a).

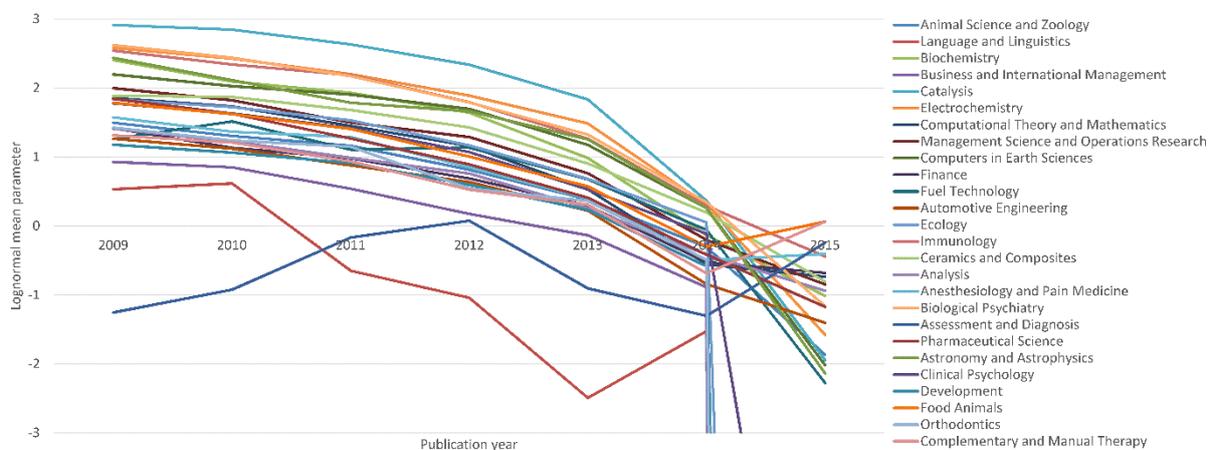

**Figure 2**. The mean parameter from the discretised lognormal distribution for each subject and year, including all uncited articles, and with 1 added to all citation counts. High negative values are excluded for clarity.

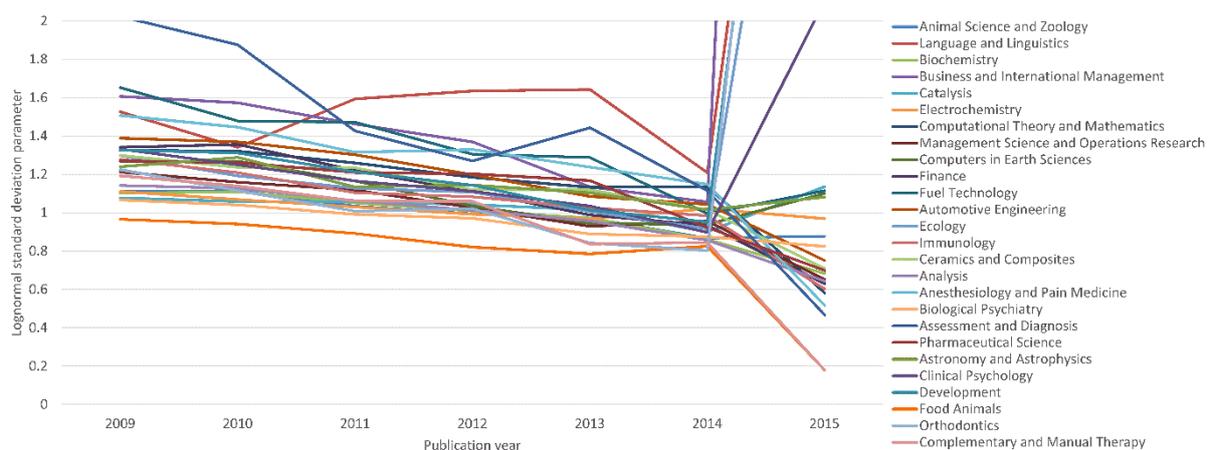

**Figure 3**. The standard deviation parameter from the discretised lognormal distribution for each subject and year, including all uncited articles, and with 1 added to all citation counts. High positive values are excluded for clarity.

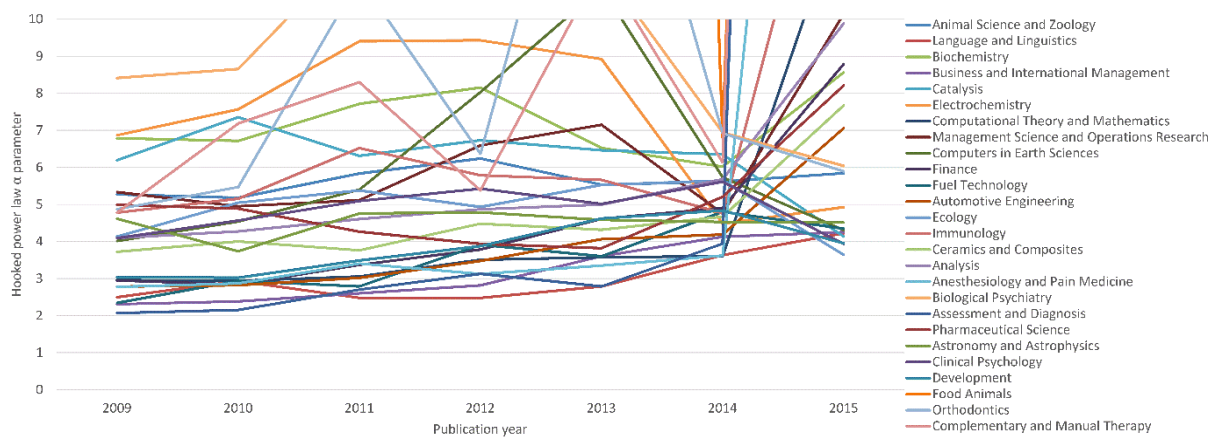

**Figure 4**. The alpha parameter from the hooked power law for each subject and year, including all uncited articles, and with 1 added to all citation counts. High positive values are excluded for clarity.

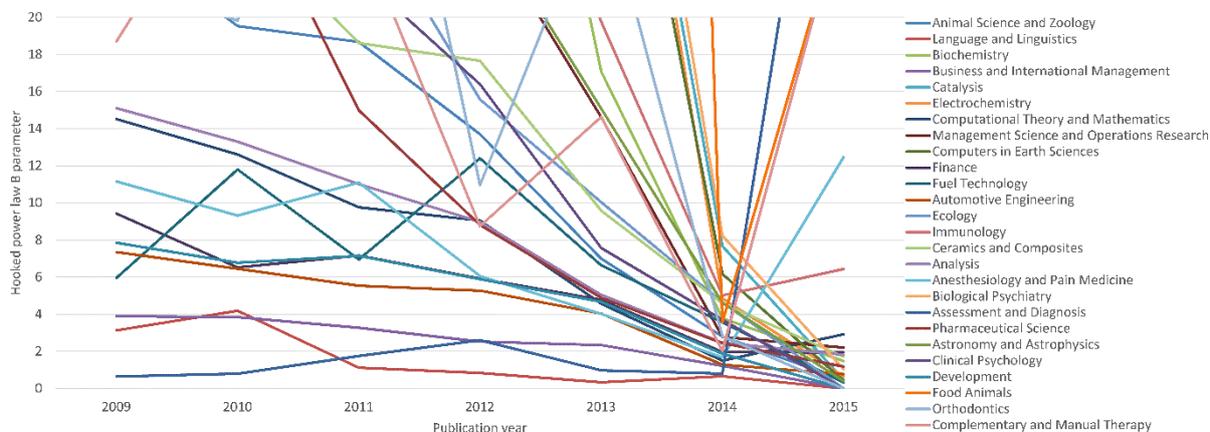

**Figure 5**. The B parameter from the hooked power law for each subject and year, including all uncited articles, and with 1 added to all citation counts. High positive values are excluded for clarity.

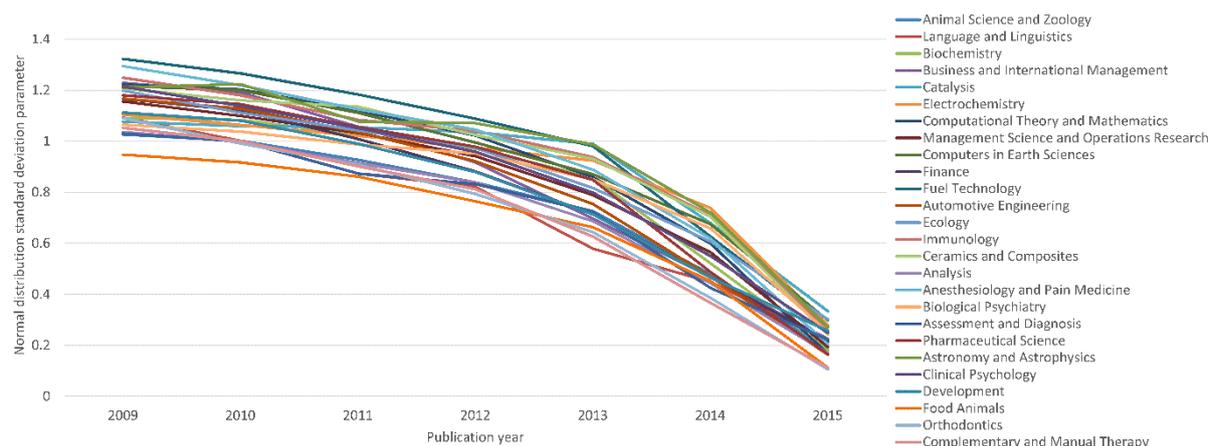

**Figure 6**. The mean parameter from the normal distribution for each subject and year, including all uncited articles, and with 1 added to all citation counts.

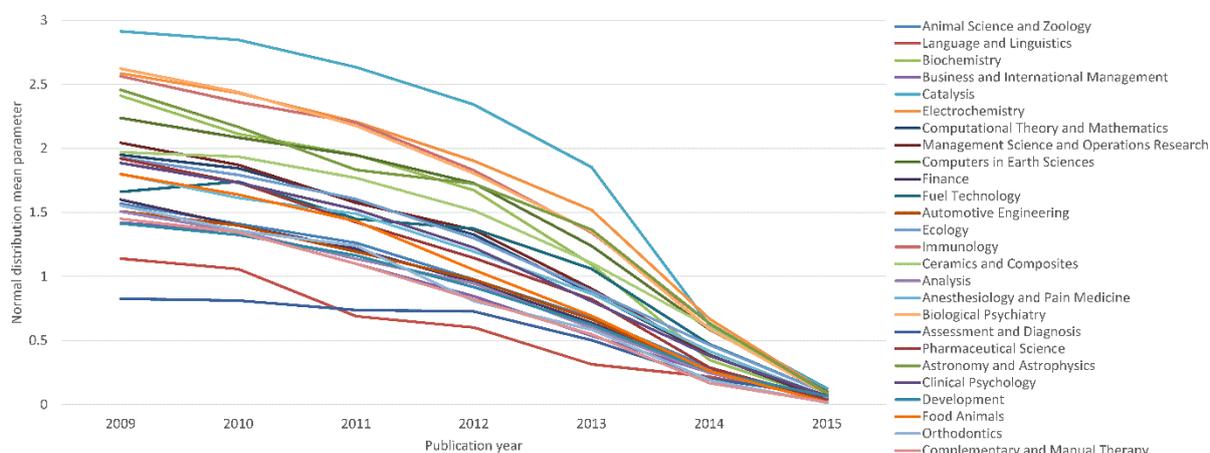

**Figure 7**. The standard deviation parameter from the normal distribution for each subject and year, including all uncited articles, and with 1 added to all citation counts.

## 7. Conclusions

The results show that, when full sets of citation data are analysed (i.e., including uncited articles), both the discretised lognormal and the hooked power law fit better for some subject areas than does the other distribution. The hooked power law fits best overall and for citation counts of up to a year old and for natural, life and medical sciences, whereas the discretised lognormal distribution tends to fit better for older articles for arts, humanities, social science, engineering and technology areas. The difference between the fits of the two distributions is not large however, and in many cases was not statistically significant despite the relatively large data sets (over 1000 articles per year in all except three subject areas). Hence it would be reasonable to apply either of the distributions to all subject areas if necessary, even those for which one was a slightly better fit than the other. The software provided can be used for hooked power law fitting.

Both the discretised lognormal and the hooked power law have problems with parameter estimates from the fitted model not being stable over time, especially in some subject areas and for recent articles, and so the standard normal distribution applied to the natural log of the citation counts plus one seems to be the best approach when accurate model parameters are needed, as in the context of regression. This conclusion is despite the normal distribution fitting method being designed for continuous data and the other two being fit directly as discrete distributions. In particular, for regression analyses of citation data from a single subject, adding 1 to the citation counts, taking the natural logarithm and then applying ordinary least squares regression is more suitable than regression based on either of the discrete models. This conclusion applies only to citation count data because the strategy has been shown not to work well for some other kinds of skewed distributions (O'Hara & Kotze, 2010).

Both of the above conclusions apply to sets of academic journal articles that are from a single subject area rather than broad collections of articles from multiple subject areas (or years). Additional research is needed to assess the extent to which the finding could also apply to multidisciplinary collections of articles that may essentially combine distributions with either different models or different model parameters. In this regard, the normal distribution finding for regression is promising because, if a normalisation process was applied to merged sets of articles (e.g., combing multiple fields after field normalisation by dividing the citation counts for each field by the mean) then this would yield non-discrete data, which would be a problem for the two discrete distributions.

13Brzezinski, M. (2015). Power laws in citation distributions: Evidence from Scopus. Scientometrics, 103(1), 213-228.

Chandy, P. R., & Williams, T. G. (1994). The impact of journals and authors on international business research: A citational analysis of JIBS articles. Journal of International Business Studies, 715-728.

Charlton, B. G., & Andras, P. (2007). Evaluating universities using simple scientometric research-output metrics: total citation counts per university for a retrospective seven-year rolling sample. *Science and Public Policy*, 34(8), 555-563.

Clauset, A., Shalizi, C. R., & Newman, M. E. (2009). Power-law distributions in empirical data. SIAM review, 51(4), 661-703.

de Solla Price, D.J. (1965). Networks of scientific papers. Science, 149, 510-515.

de Solla Price, D.J. (1976). A general theory of bibliometric and other cumulative advantage processes. Journal of the American Society for Information Science, 292-306.

Didegah, F., & Thelwall, M. (2013). Which factors help authors produce the highest impact research? Collaboration, journal and document properties. Journal of Informetrics, 7(4), 861-873.

Dorogovtsev, S. N., Mendes, J. F. F., & Samukhin, A. N. (2000). Structure of growing networks with preferential linking. Physical review letters, 85(21), 4633.

Eom, Y. H., & Fortunato, S. (2011). Characterizing and modeling citation dynamics. PLoS One, 6(9), e24926.

Evans, T.S., Hopkins, N. & Kaube, B.S. (2012). Universality of performance indicators based on citation and reference counts. Scientometrics, 93, 473-495.

Fairclough, R., & Thelwall, M. (2015a). More precise methods for national research citation impact comparisons. *Journal of Informetrics*, 9(4), 895–906.

Fairclough, R. & Thelwall, M. (2015b). National research impact indicators from Mendeley readers. Journal of Informetrics, 9(4), 845–859.

Garanina, O. S., & Romanovsky, M. Y. (2015). Citation distribution of individual scientist: Approximations of Stretch Exponential Distribution with Power Law Tails. In Salah, A.A., Y. Tonta, A.A. Akdag Salah, C. Sugimoto, U. Al (Eds.) Proceedings of ISSI 2015. Istanbul, Turkey: Bogaziçi University Printhouse (pp. 272-277).

Gazni, A., & Didegah, F. (2011). Investigating different types of research collaboration and citation impact: a case study of Harvard University's publications. Scientometrics, 87(2), 251-265.

Glänzel, W., Schubert, A., & Czerwon, H. J. (1999). A bibliometric analysis of international scientific cooperation of the European Union (1985–1995). Scientometrics, 45(2), 185-202.

Hanssen, T. E. S., & Jørgensen, F. (2015). The value of experience in research. Journal of Informetrics, 9(1), 16-24.

Harnad, S., & Brody, T. (2004). Comparing the impact of open access (OA) vs. non-OA articles in the same journals. D-lib Magazine, 10(6). http://www.dlib.org/dlib/june04/harnad/06harnad.html

Hellqvist, B. (2010). Referencing in the humanities and its implications for citation analysis, Journal of the American Society for Information Science and Technology, 61(2), 310-318.

Hicks, D. (1999). The difficulty of achieving full coverage of international social science literature and the bibliometric consequences. Scientometrics, 44(2), 193-215.

## 9. Appendix

The programs, data, and fitting parameters and statistics are in the online appendix at Figshare: DOI: 10.6084/m9.figshare.2056707

**Table 3**. Summary statistics about the subjects analysed.

| Publication year | 2009 | 2009 | 2009 | 2010 | 2010 | 2010 | 2011 | 2011 | 2011 | 2012 | 2012 | 2012 | 2013 | 2013 | 2013 | 2014 | 2014 | 2014 | 2015 | 2015 | 2015 |
|---|---|---|---|---|---|---|---|---|---|---|---|---|---|---|---|---|---|---|---|---|---|
| Subject | Mean | Max. | N | Mean | Max. | N | Mean | Max. | N | Mean | Max. | N | Mean | Max. | N | Mean | Max. | N | Mean | Max. | N |
| Animal Science & Zoology | 7.23 | 433 | 9968 | 5.61 | 182 | 9994 | 4.56 | 557 | 9761 | 2.83 | 119 | 9974 | 1.66 | 88 | 9995 | 0.52 | 31 | 9990 | 0.07 | 11 | 6221 |
| Language and Linguistics | 3.69 | 1016 | 7900 | 2.68 | 129 | 9992 | 1.8 | 89 | 9820 | 1.55 | 88 | 9982 | 0.68 | 101 | 9984 | 0.32 | 20 | 7775 | 0.07 | 7 | 3308 |
| Biochemistry | 18.96 | 652 | 9971 | 14.28 | 544 | 9922 | 11.06 | 893 | 9920 | 7.85 | 641 | 9918 | 3.58 | 194 | 9883 | 0.66 | 31 | 9914 | 0.07 | 7 | 9935 |
| Business & Int. Man. | 4.59 | 98 | 9994 | 4.24 | 1259 | 9996 | 2.38 | 96 | 8418 | 1.92 | 217 | 9997 | 0.94 | 58 | 9976 | 0.32 | 37 | 8567 | 0.07 | 12 | 3477 |
| Catalysis | 32.61 | 1621 | 9559 | 28.26 | 868 | 9628 | 22.89 | 713 | 9648 | 16.37 | 496 | 9706 | 8.97 | 99 | 9729 | 1.35 | 35 | 9620 | 0.2 | 34 | 7699 |
| Electrochemistry | 21.75 | 698 | 8748 | 18.13 | 807 | 9805 | 13.52 | 372 | 9768 | 9.55 | 291 | 9803 | 6 | 308 | 9880 | 1.66 | 202 | 9274 | 0.14 | 7 | 4099 |
| Comp. Theory & Math. | 8.74 | 3734 | 9735 | 5.38 | 1023 | 9703 | 4.04 | 94 | 9799 | 3.62 | 297 | 9524 | 1.42 | 97 | 9600 | 0.6 | 144 | 8672 | 0.04 | 4 | 4059 |
| Management Science & OR | 8.21 | 390 | 9235 | 7.7 | 4302 | 8618 | 5.45 | 194 | 8215 | 4.32 | 120 | 8066 | 2.08 | 61 | 8660 | 0.64 | 28 | 7867 | 0.07 | 3 | 2945 |
| Computers in Earth Sci. | 13.28 | 564 | 1794 | 10.26 | 98 | 1596 | 8.24 | 306 | 2047 | 7.05 | 276 | 2133 | 3.74 | 100 | 1892 | 1.05 | 33 | 2681 | 0.13 | 4 | 922 |
| Finance | 8.15 | 1439 | 7875 | 6.34 | 458 | 8151 | 3.75 | 246 | 9646 | 2.97 | 119 | 7936 | 1.48 | 64 | 7918 | 0.43 | 17 | 7669 | 0.1 | 9 | 2485 |
| Fuel Technology | 5.78 | 522 | 9964 | 5.68 | 490 | 9972 | 4.77 | 348 | 9980 | 3.68 | 343 | 9973 | 1.92 | 101 | 9987 | 0.71 | 23 | 9965 | 0.13 | 7 | 4248 |
| Automotive Engineering | 3.81 | 194 | 8012 | 3.78 | 224 | 6972 | 2.25 | 92 | 9711 | 1.69 | 69 | 8505 | 0.84 | 34 | 9537 | 0.29 | 35 | 5671 | 0.04 | 4 | 1535 |
| Ecology | 11.82 | 2573 | 9410 | 9.98 | 439 | 9398 | 7.18 | 192 | 9408 | 4.99 | 159 | 9350 | 2.5 | 68 | 9254 | 0.87 | 35 | 9081 | 0.15 | 38 | 5893 |
| Immunology | 23.29 | 1493 | 9224 | 18.9 | 859 | 9640 | 14.55 | 638 | 9608 | 9.32 | 244 | 9659 | 5.03 | 394 | 9712 | 1.34 | 68 | 9616 | 0.13 | 5 | 6667 |
| Ceramics & Composites | 9.49 | 388 | 8965 | 10.19 | 857 | 8033 | 5.85 | 98 | 8545 | 4.48 | 299 | 8596 | 3.59 | 168 | 8204 | 1.19 | 59 | 8635 | 0.11 | 12 | 6231 |
| Analysis | 6.92 | 736 | 8286 | 5.38 | 212 | 8074 | 3.84 | 91 | 8616 | 2.81 | 136 | 9437 | 1.38 | 48 | 9778 | 0.41 | 24 | 8958 | 0.05 | 4 | 3412 |
| Anesth. & Pain Medicine | 10.32 | 742 | 9980 | 7.83 | 383 | 9976 | 5.99 | 486 | 9665 | 4.13 | 134 | 9977 | 2.27 | 77 | 9979 | 0.75 | 46 | 7525 | 0.07 | 4 | 3656 |
| Biological Psychiatry | 20.94 | 804 | 3559 | 16.7 | 663 | 3698 | 12.58 | 512 | 3832 | 8.03 | 144 | 4343 | 4.57 | 128 | 5233 | 1.22 | 24 | 5717 | 0.13 | 7 | 1798 |
| Assessment & Diagnosis | 3.01 | 92 | 735 | 2.67 | 151 | 817 | 1.63 | 45 | 684 | 1.67 | 26 | 675 | 1 | 26 | 669 | 0.29 | 8 | 633 | 0.06 | 2 | 217 |
| Pharmaceutical Sci. | 10.85 | 437 | 9472 | 8.73 | 510 | 9564 | 5.85 | 324 | 9471 | 4.19 | 240 | 9501 | 2.5 | 252 | 9628 | 0.53 | 24 | 9557 | 0.05 | 6 | 6400 |
| Astron. & Astrophysics | 16.83 | 2983 | 9988 | 13.16 | 1409 | 9968 | 8.92 | 3267 | 9996 | 6.49 | 567 | 9990 | 4.47 | 712 | 9993 | 1.38 | 75 | 9980 | 0.14 | 19 | 3153 |
| Clinical Psychology | 11.33 | 318 | 9999 | 9.34 | 481 | 9999 | 6.77 | 220 | 9994 | 4.4 | 141 | 9907 | 2.24 | 84 | 9958 | 0.69 | 35 | 8725 | 0.1 | 14 | 4934 |
| Development | 6.25 | 99 | 5835 | 5.64 | 337 | 6172 | 4.22 | 287 | 6619 | 2.71 | 73 | 6830 | 1.48 | 50 | 6903 | 0.44 | 19 | 6477 | 0.11 | 20 | 2301 |
| Food Animals | 7.86 | 117 | 1643 | 6.77 | 87 | 1814 | 4.96 | 65 | 1759 | 2.9 | 35 | 2082 | 1.59 | 32 | 1837 | 0.49 | 9 | 1771 | 0.03 | 1 | 147 |
| Orthodontics | 6.39 | 83 | 1148 | 4.61 | 72 | 1246 | 3.52 | 52 | 1242 | 1.84 | 31 | 1278 | 1.12 | 17 | 1223 | 0.27 | 7 | 1070 | 0.02 | 2 | 311 |
| Comp. & Man. Therapy | 5.82 | 54 | 331 | 4.82 | 37 | 372 | 3.22 | 29 | 389 | 2.36 | 28 | 386 | 1.14 | 12 | 442 | 0.31 | 10 | 476 | 0.02 | 1 | 124 |